\documentclass[reprint,reprintnumbers,amsmath,amssymb]{revtex4-1}

\usepackage{graphicx}
\usepackage{dcolumn}
\usepackage{bm}

\usepackage{epstopdf} 
\usepackage{upgreek}
\usepackage{color}
\usepackage[dvipsnames]{xcolor}

\newcommand{\D}{\text{D}}
\newcommand{\G}{\text{G}}
\newcommand{\TG}{\text{TG}}
\newcommand{\BG}{\text{BG}}

\begin{document}

\title{Weak Localization and Weak Antilocalization in Double-Gate a-InGaZnO Thin-Film Transistors}

\author{Wei-Hsiang Wang}
\affiliation{Department of Physics, National Taiwan Normal University, Taipei 116, Taiwan}
\author{Elica Heredia}
\affiliation{Department of Physics, National Taiwan Normal University, Taipei 116, Taiwan}
\author{Syue-Ru Lyu}
\affiliation{Department of Physics, National Taiwan Normal University, Taipei 116, Taiwan}
\author{Shu-Hao Liu}
\affiliation{Department of Physics, National Taiwan Normal University, Taipei 116, Taiwan}
\author{Po-Yung Liao}
\affiliation{Department of Physics, National Sun Yat-Sen University, Kaohsiung 804, Taiwan}
\author{Ting-Chang Chang}
\affiliation{Department of Physics, National Sun Yat-Sen University, Kaohsiung 804, Taiwan}
\author{Pei-hsun Jiang}
\altaffiliation{E-mail: pjiang@ntnu.edu.tw}
\affiliation{Department of Physics, National Taiwan Normal University, Taipei 116, Taiwan}

\begin{abstract}
We demonstrate manipulation of quantum interference by controlling the competitions between weak localization (WL) and weak antilocalization (WAL) via variation of the gate voltages of double-gate amorphous InGaZnO thin-film transistors. Our study unveils the full profile of an intriguing universal dependence of the respective WL and WAL contributions on the channel conductivity. This universality is discovered to be robust against interface disorder.
\end{abstract}

\maketitle

Measurements of magnetoconductivity are known to be a powerful tool to study spin effects. \textit{Weak localization} (WL) refers to constructive quantum interference of coherently back-scattered conduction electrons, which leads to a suppressed conductivity \cite{Altshuler1985}. \textit{Weak antilocalization} (WAL), on the other hand, refers to destructive interference due to rotated spins of the waves in the opposite direction in the presence of spin--orbit coupling (SOC), leading to an enhanced conductivity \cite{Hikami1980}. WL and WAL have recently been explored in doped ZnO films and nanowires because of potential applications in nanoelectronics and spintronics \cite{Ozgur2005,Shinozaki2007,Thompson2009,Xu2010,Shinozaki2013,Chiu2013,Yabuta2014,Kulbachinskii2015}. Crossovers from WL to WAL have been observed in magnetoconductivity of InZnO films via temperature variation \cite{Shinozaki2013,Yabuta2014}. However, gate-controlled quantum interference in ZnO or its doped form was not reported until very recently, when competitions between WL and WAL were discovered in single-gate a-IGZO TFTs via electric gating in our previous research \cite{Wang2017}, 
and an intriguing universal dependence of the respective WL and WAL contributions on the channel conductivity was observed \cite{Wang2017a}. In this paper, more information about this universal dependence is revealed by quantum magnetotransport measurements on double-gate a-IGZO TFTs with higher conductivity. This universal dependence is found to be robust against interface disorder, and therefore the conductivity can reliably reveal the strength of the SOC effect tuned by variation of a single or multiple gate voltages of spintronic devices.

\begin{figure}[!t]
\centering
\includegraphics[width=3.4in]{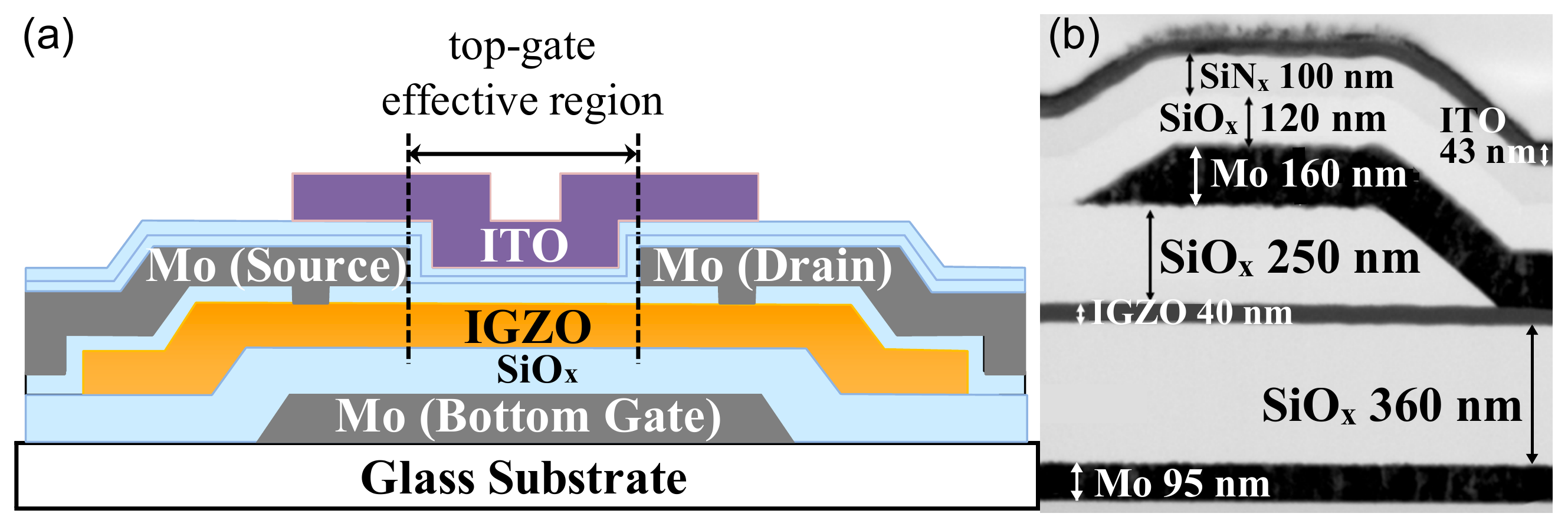}
\caption{\label{fig:fig1}
(a) Schematic drawing and (b) the TEM image of the cross-sectional view of the double-gate a-IGZO thin-film transistor.
}
\end{figure}

The a-IGZO TFTs with an inverted staggered via-contact structure are fabricated on glass as shown in Fig.~\ref{fig:fig1}. First, a Mo film was sputtered as bottom gate electrodes, followed by a SiO$_x$ gate insulating layer deposited by plasma-enhanced chemical vapor deposition (PECVD). Next, a 40-nm-thick a-IGZO channel layer was sputtered at room temperature with a targeted In$_2$O$_3$:Ga$_2$O$_3$:ZnO atomic ratio of 1:1:1. A SiO$_x$ etching stop layer was then deposited by PECVD at 200 $^\circ$C. The via-contact-type source and drain electrodes were formed by sputtering Mo. A SiO$_x$/SiN$_x$ film was deposited as the passivation layer using PECVD. Indium-tin-oxide (ITO) electrodes were then formed as the top gates. Finally, the device was annealed at 240 $^\circ$C in an atmospheric oven.

\begin{figure}[!t]
\centering
\includegraphics[width=3.6in]{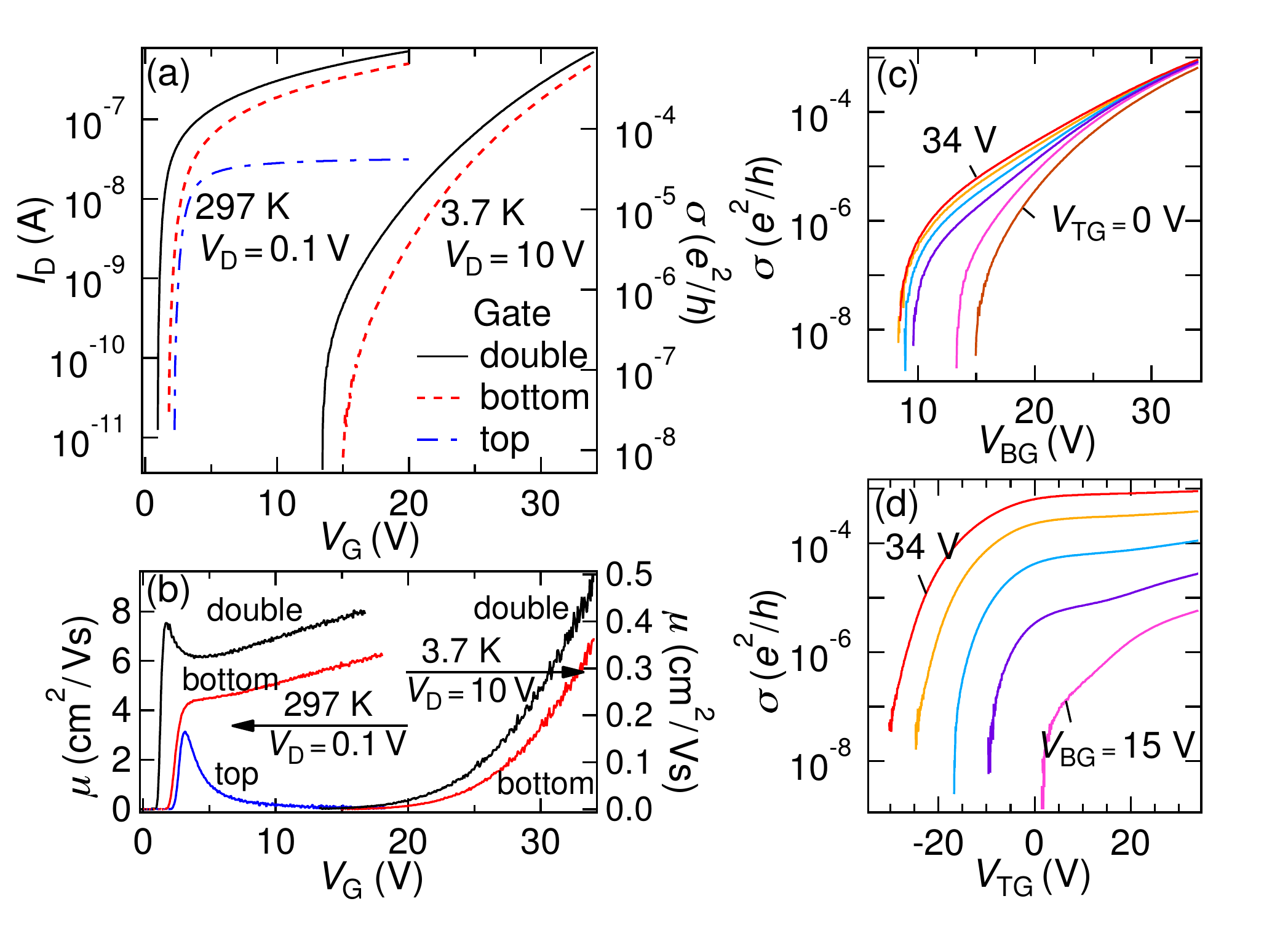}
\caption{\label{fig:fig2}
(a) Drain currents ($I_\D$) as functions of the bottom-, top-, or double-gate voltages. Room-temperature data are taken with a drain voltage ($V_\D$) of 0.1 V, whereas 3.7-K data are taken with $V_\D=10$ V. The corresponding 2D conductivity ($\sigma$) is appended on the right axis. (b) Respective field-effect mobilities ($\mu$) as functions of the gate voltages. $\mu=dI_\D/dV_\G\cdot L/(WC_{\text{ins}}V_\D)$, where $C_{\text{ins}}$ is the insulating-layer capacitance measured as functions of the bottom-, top-, or double-gate voltages, respectively. $\mu$ curves at 3.7 K are appended to the right axis.
(c) $\sigma$ at 3.7 K with $V_\D=10$ V as functions of the bottom-gate voltage ($V_\BG$) at various top-gate voltages ($V_\TG$) of 0, 10, 20, 25, 30, and 34 V, respectively from bottom to top. (d) $\sigma$ at 3.7 K with $V_\D=10$ V as functions of $V_\TG$ at various $V_\BG$ of 15, 20, 25, 30, and 34 V, respectively from bottom to top.
}
\end{figure}

Electrical measurements were performed on several double-gate a-IGZO TFTs. 
Low-temperature measurements were conducted with the device mounted in a cryogenic system equipped with a superconducting magnet. Connections to the electrodes are made via wire bonding. 
The data presented in this paper are collected from a representative sample with a channel width ($W$) of 20 $\upmu$m and a channel length ($L$) of 10 $\upmu$m. Fig.~\ref{fig:fig2}a shows $I_\D$ under a bottom-gate voltage ($V_\BG$) sweep with a grounded top gate, a top-gate voltage ($V_\TG$) sweep with a grounded bottom gate, and a double-gate voltage sweep with $V_\BG=V_\TG$, respectively. Their respective field-effect mobilities ($\mu$) are displayed in Fig.~\ref{fig:fig2}b. The three curves on the left in Fig.~\ref{fig:fig2}a are taken at room temperature with $V_\D = 0.1$ V. For the $V_\BG$ sweep with $V_\TG =0$, the threshold gate voltage is 1.9 V, the subthreshold swing is 0.18 V/decade, and $\mu\sim4.8$ cm$^2$/Vs and $I_\D \sim 0.13$ $\upmu$A when entering the linear regime. For the double-gate voltage sweep, $I_\D$ behaves similarly to that under the $V_\BG$ sweep, but $I_\D$ and $\mu$ at the beginning of the linear regime increase to $\sim$0.23 $\upmu$A and $\sim$6.4 cm$^2$/Vs, respectively.  If instead $V_\TG$ is swept with $V_\BG=0$,  $I_\D$ is much smaller than the other two sweeps, and stays almost constant  for $V_\TG > 10$ V.  The small constant  $I_\D$ can be interpreted as the electron-injection-induced diffusion current due to the screening of top-gate electric field by the redundant source and drain electrodes \cite{Tsai2015}, as illustrated in Fig.~\ref{fig:fig1}, leading to an effective  length of only $\sim$3/4 of the channel length \cite{Liao2016}.

At a low temperature ($T$) of 3.7 K, the threshold gate voltage is seriously enhanced, and $I_\D$ is not detectable with $V_\D = 0.1$ V for any gate voltages less than 35 V. Therefore, a larger $V_\D$ of 10 V is supplied for all 3.7-K measurements to give detectable $I_\D$ \cite{VD}. At 3.7 K, $I_\D$ increases much more slowly with the gate voltages, as shown in Fig.~\ref{fig:fig2}a, and is not even detectable at any $V_\TG$ below 35 V when $V_\BG=0$ because of the extremely low diffusion rate of the carriers. $\mu$ is only $\sim$0.5 cm$^2$/Vs when $V_\TG=V_\BG=34$ V, and remains higher under the double-gate voltage sweep than under the $V_\BG$ sweep. 
The seriously suppressed $\mu$ at low $T$ can be interpreted in the percolation model for  amorphous oxide semiconductor TFTs \cite{Hosono2006, Lee2011}. Shown on the right axis of Fig.~\ref{fig:fig2}a is the 2D conductivity ($\sigma$) given by $I_\D/V_\D\cdot L/W$, and is expressed in unit of $e^2/h$, where $h$ is the Planck's constant.

The conductivity ($\sigma$) at 3.7 K was further measured under various combinations of gate voltages. Fig.~\ref{fig:fig2}c presents $\sigma$ vs $V_\BG$ at various fixed $V_\TG$, whereas Fig.~\ref{fig:fig2}d shows $\sigma$ vs $V_\TG$ at various fixed $V_\BG$. It can be seen that the two gates control the carriers in the channel in very different manners. In Fig.~\ref{fig:fig2}c, all $\sigma$ curves of different $V_\TG$ approach an ``asymptote'' that passes through $\sim$$10^{-3} e^2/h$ at $V_\BG= 34$ V. However, $V_\TG= 34$ V, as shown in Fig.~\ref{fig:fig2}d, gives much smaller values of $\sigma$ at various smaller $V_\BG$ until $V_\BG$ reaches 34 V to give $\sigma \sim 10^{-3} e^2/h$. These results reflect the limited effectiveness of $V_\TG$.

\begin{figure}[!t]
\centering
\includegraphics[width=3.5in]{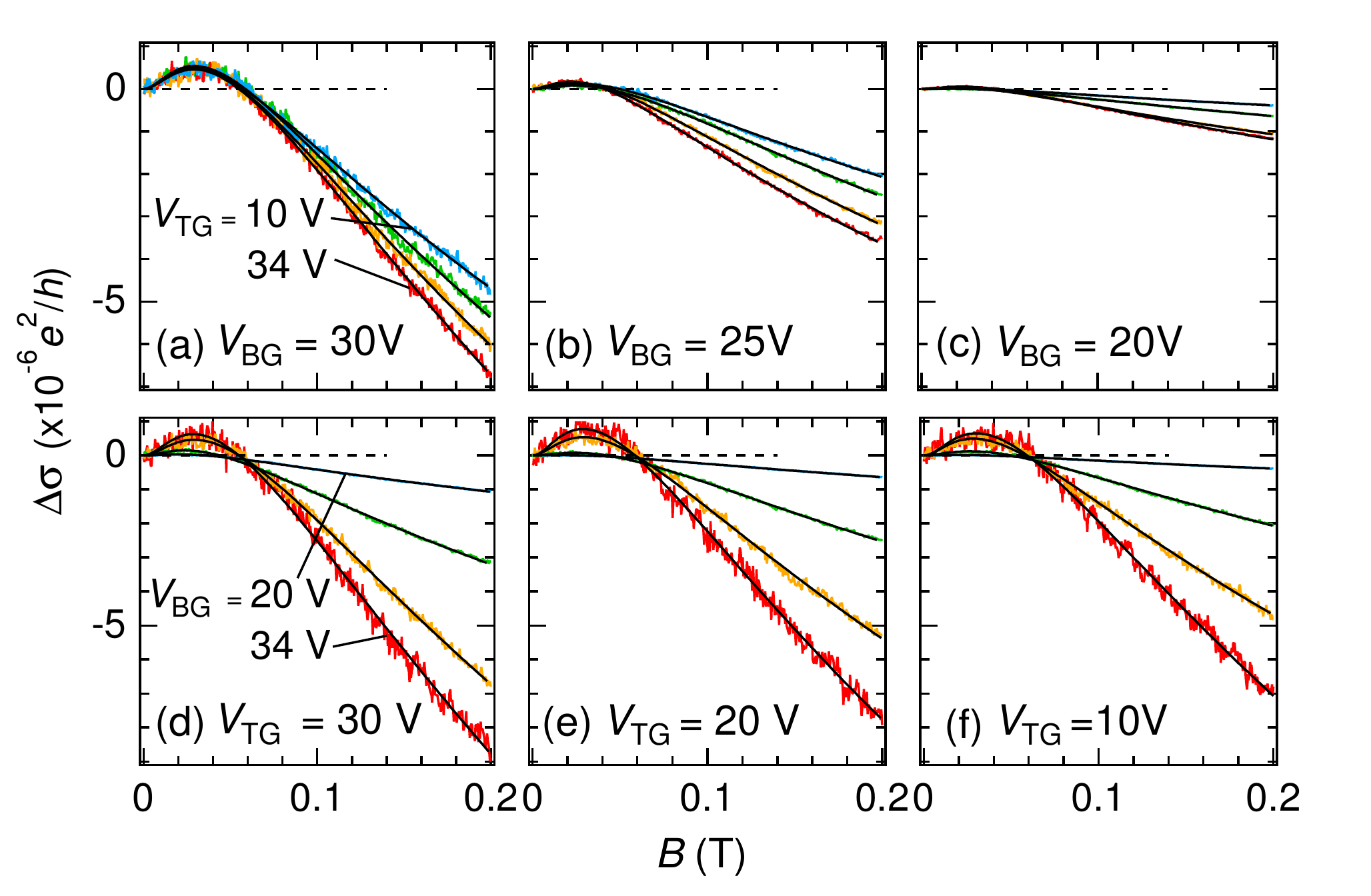}
\caption{\label{fig:fig3}
$\Delta \sigma$ at 3.7 K as functions of the magnetic field ($B$) (a)--(c) at fixed 
$V_\BG$ (30, 25, or 20 V) with various $V_\TG$ of 10, 20, 30, and 34 V, respectively from top to bottom, and (d)--(f) at fixed $V_\TG$ (30, 20, or 10 V) with various $V_\BG$ of 20, 25, 30, and 34 V, respectively from top to bottom. Theoretical fits (Eq.~\ref{eq:1}) are shown as solid smooth lines.
}
\end{figure}

The quantum magnetotransport are investigated at 3.7 K with magnetic field ($B$) perpendicular to the channel plane. The representative curves of $\Delta \sigma(B) (\equiv \sigma(B)-\sigma(0)$) are shown in Fig.~\ref{fig:fig3}. Figs.~\ref{fig:fig3}a--\ref{fig:fig3}c display $\Delta \sigma(B)$ at respective fixed $V_\BG$ with various $V_\TG$ from 10 V to 34 V, whereas Figs.~\ref{fig:fig3}e--\ref{fig:fig3}f display $\Delta \sigma$ at respective fixed $V_\TG$ with various $V_\BG$ from 20 V to 34 V. In general, $\Delta \sigma$ curves with higher gate voltages increase with $B$ up to $B \sim 0.037$ T, demonstrating a WL signature, but then bend over to WAL and decrease substantially at higher $B$. The WL and WAL features are suppressed as either gate voltage is decreased, but with different effectiveness. Fig.~\ref{fig:fig3} shows that varying $V_\BG$ substantially changes the magnetotransport of the device, whereas varying $V_\TG$ only tunes it mildly.

To assess the respective contributions of WL and WAL for each $\Delta \sigma(B)$ curve, 
we use the two-component Hikami--Larkin--Nagaoka (HLN) theory for the magnetoconductivity of a 2D system in the limit of strong SOC \cite{Hikami1980,Lu2011,Lu2011a}: 
\begin{equation}\label{eq:1}
\Delta\sigma(B)=A\sum_{i=0, 1}\frac{\alpha_i e^2}{\pi h}\Bigg[\Psi \Bigg(\frac{\ell_B^2}{\ell_{\phi i}^2}+\frac{1}{2} \Bigg) - \ln \Bigg( \frac{\ell_B^2}{\ell_{\phi i}^2} \Bigg)\Bigg],
\end{equation}
where $\Psi$ is the digamma function, $\ell_B \equiv \sqrt{\hbar/(4e|B|)}$ is half the magnetic length, the prefactors $\alpha_0$ and $\alpha_1$ stand for the weights of WL and WAL, respectively, and $\ell_{\phi i}$ is the corresponding phase coherence length. The original two-component HLN equation was proposed for topological insulator thin films  \cite{Lu2011,Lu2011a}. Owing to the percolation conduction in a-IGZO \cite{Hosono2006}, it is believed that the effective $L$ is much larger and the effective $W$ is much smaller than the values of the channel dimensions\cite{Shinozaki2011,Kamiya2009}. This implies that the real conductivity is much larger than $\sigma \equiv I_\D/V_\D\cdot L/W$. Therefore, we add a small coefficient $A$ to the original equation \cite{Lu2011,Lu2011a} to address this issue. The magnitudes of $\sigma$ of the double-gate a-IGZO TFTs are found to be $\sim$3 times larger than those of the single-gate a-IGZO TFTs from our previous measurements \cite{Wang2017, Wang2017a}, so $A$ is set to $3 \times 10^{-4}$ in this work to roughly compensate the difference in sample quality \cite{A}. Eq.~\ref{eq:1} provides excellent fits to all the curves in Fig.~\ref{fig:fig3}, which are shown as solid smooth lines.

\begin{figure}[!t]
\centering
\includegraphics[width=3.8in]{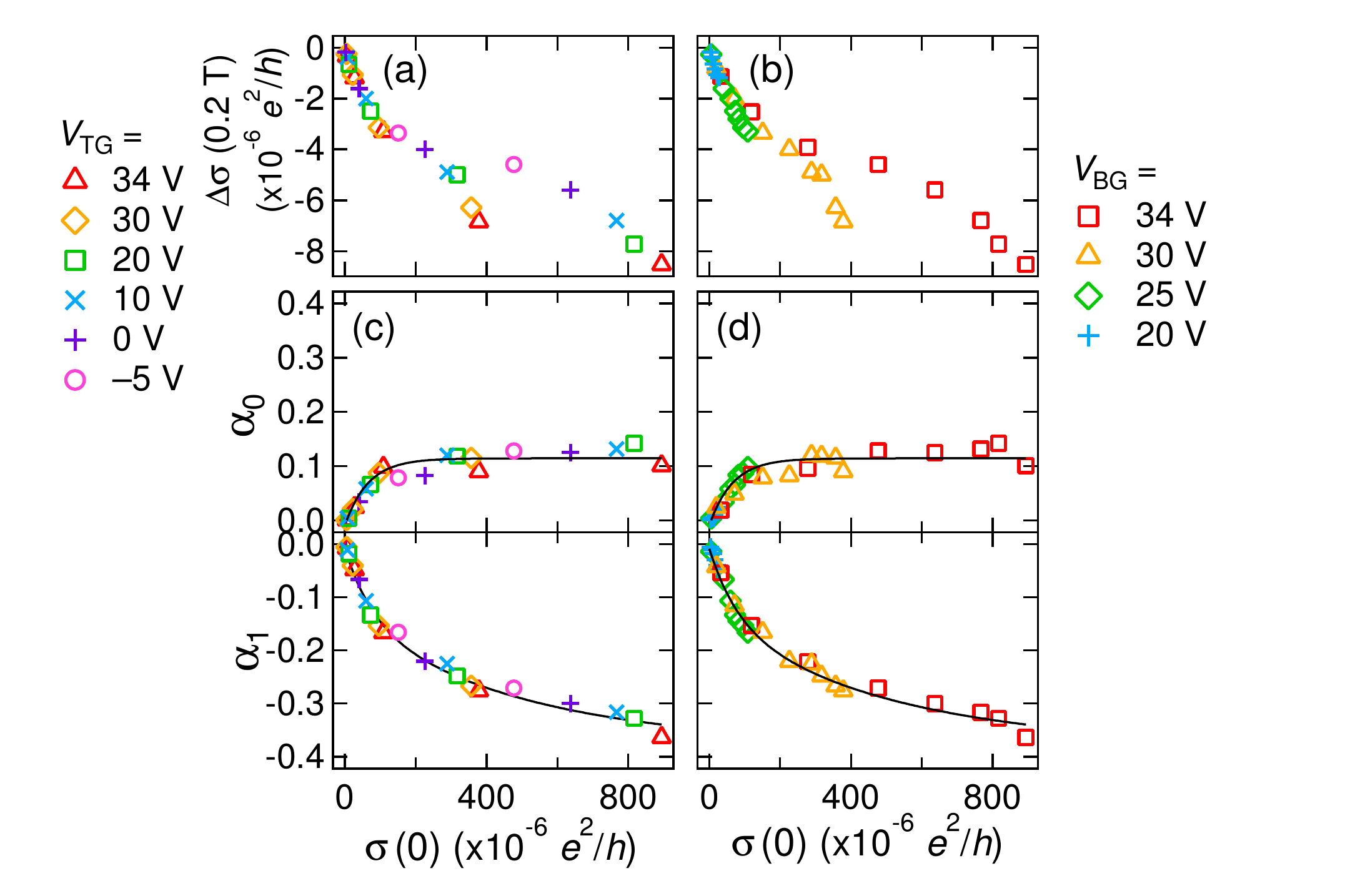}
\caption{\label{fig:fig4}
$\Delta \sigma(0.2 \text{ T})$ vs $\sigma(0)$ (a) at various fixed $V_\TG$, and (b) at various fixed $V_\BG$. There are duplicate data points in (a) and (b); for example, ``\textcolor{violet}{+}" at $\sigma(0)=637\times10^{-6} e^2/h$ in (a) and ``\textcolor{red}{$\Box$}" at the same $\sigma(0)$ in (b) are the same data point at $V_\TG=0$ and $V_\BG=34$ V. $\alpha_0$ (the prefactor for WL) and $\alpha_1$ (the prefactor for WAL) obtained from the theoretical fits (Eq.~\ref{eq:1}) at various gate voltages used in (a) and (b) are shown in (c) and (d), respectively. Eye-guiding lines of the universal curves are shown as solid smooth lines.
}
\end{figure}

Measurements on the double-gate a-IGZO TFTs with higher $\sigma(0)$ ($\sigma$ at zero $B$) allow us to investigate the evolution of the WL and WAL competitions from zero- to high-conductivity regimes. In our previous experiment on single-gate a-IGZO TFTs \cite{Wang2017a}, only low-conductivity transport ($\sigma(0)\lesssim 200\times10^{-6}e^2/h$) could be studied. Figs.~\ref{fig:fig4}a and \ref{fig:fig4}b display $\Delta \sigma(0.2 \text{ T})$ vs $\sigma(0)$ to demonstrate that different gate operations may give different $\Delta \sigma$ at the same $\sigma(0)$ and $B$ when $\sigma(0)$ is large ($> 200 \times 10^{-6}e^2/h$ for the case of $B=0.2$ T). Fig.~\ref{fig:fig4}a shows $\Delta \sigma (0.2 \text{ T})$ at various fixed $V_\TG$ as functions of $\sigma(0)$ achieved by varying $V_\BG$, whereas Fig.~\ref{fig:fig4}b shows $\Delta \sigma (0.2 \text{ T})$ at various fixed $V_\BG$ as functions of $\sigma(0)$ achieved by varying $V_\TG$. The values of $\alpha_0$ and $\alpha_1$ obtained by fitting experimental $\Delta \sigma(B)$ curves at respective gate voltages to Eq.~\ref{eq:1} are plotted against $\sigma(0)$ in Figs.~\ref{fig:fig4}c and \ref{fig:fig4}d \cite{uncertainty}, and each of them surprisingly collapses onto a universal curve despite the distinct patterns of $\Delta \sigma(0.2 \text{ T})$ vs $\sigma(0)$ obtained from different gate operations shown in Figs.~\ref{fig:fig4}a and \ref{fig:fig4}b. $\alpha_0$ increases rapidly from 0 to $\sim$0.1 with increasing $\sigma(0)$, but then stays around 0.1 after the transport passes the threshold.  $|\alpha_1|$ grows substantially with increasing $\sigma(0)$, and then slows down to approach $\sim$0.35 when $\sigma(0)$ gets close to $\sim$$10^{-3} e^2/h$. It is noticed that the shape of the universal curve may vary among devices with different structure fabrications. The single-gate a-IGZO TFTs in our previous experiment \cite{Wang2017a}, for example, never entered the linear regime within the $V_\G$ range studied, and therefore its partial universal curve of $\alpha_1$ vs $\sigma(0)$ did not show any changes in the slope.

The values of $\ell_{\phi i}$ obtained by fitting the experimental $\Delta \sigma(B)$ to Eq.~\ref{eq:1} fluctuate mildly near 0.1 $\upmu$m as $V_\BG$ or $V_\TG$ is varied. A closer look at the dependence of $\ell_{\phi i}$ on $|V_\BG- V_\TG|$ reveals that $\ell_{\phi 0}$ decreases from $\sim$0.14 $\upmu$m to $\sim$0.11 $\upmu$m and $\ell_{\phi 1}$ decreases from $\sim$0.10 $\upmu$m to $\sim$0.08 $\upmu$m as $|V_\BG- V_\TG|$ is increased from 0 to 54 V. This implies that the transport suffers stronger disorder scattering at larger vertical electric fields. In a temperature-dependence measurement, on the other hand, $\ell_{\phi i}$ is found to decrease by more than a half when $T$ is increased beyond 50 K, which weakens the WL and WAL effects. The dependence of $\alpha_i$ on $\sigma(0)$ also evolves with $T$ \cite{Wang2017}. To better observe WL and WAL, $\ell_{\phi i}$ should be much larger than the  
elastic scattering length ($\ell_{\text{e}}$) and than the spin--orbit scattering length ($\ell_{\text{SO}}$) (i.e., the strong SOC limit). Analyses on our data at 3.7 K using the original HLN equation without the assumption of strong SOC \cite{Hikami1980, Maekawa1981, Fang2016} reveal that $(\ell_{\phi i}/\ell_{\text{SO}})^2 \sim (\ell_{\phi i}/\ell_{\text{e}})^2$ ranges from $\sim$50 to $\sim$160, which satisfies the conditions for Eq.~\ref{eq:1} to be valid, and for the SOC effects to dominate for potential applications in spintronics.

Gate-voltage-controlled competitions between WL and WAL were also discovered in topological insulator thin films \cite{Lang2013}, and their $\alpha_0$ and $\alpha_1$ were theoretically shown to be determined by the ratio of the gap opening at the Dirac point to the Fermi energy \cite{Lu2011,Lu2011a}. The universal dependence of the competing WL and WAL on $\sigma(0)$ observed in a-IGZO TFT may also find its explanation in the relation between the channel conductivity and the gate-controlled Fermi-level position relative to the band structure, but this is yet to be confirmed by theoretical investigations. The a-IGZO TFTs can be used to tune the SOC effect to a desired weight via electric gating by monitoring $\sigma(0)$ thanks to the universal dependence of $\alpha_i$ on $\sigma(0)$. More research is required for future realization of ZnO-based spin transistors \cite{Datta1990} or other spintronic devices.

The double-gate structure offers a better control of the channel potential \cite{Munzenrieder2013,Zhang2001}, and thus allows a more comprehensive study. When $V_\BG=V_\TG$, the vertical electric field becomes minimal, reducing the interface roughness scattering \cite{Wong1999}. This explains why $V_\BG=V_\TG$ gives the highest $\mu$ as shown in Fig.~\ref{fig:fig2}b  and the longest $\ell_{\phi i}$ mentioned previously. However, the universal curves of  $\alpha_i$ vs $\sigma(0)$ shown in Figs.~\ref{fig:fig4}c and \ref{fig:fig4}d are robust at any $|V_\BG- V_\TG|$ regardless of the existence of interface disorder.  This implies that, although WL, WAL, and $\sigma(0)$ are all known to be affected by disorder \cite{Shinozaki2013,Yabuta2014}, the affected $\sigma(0)$ alone is somehow sufficient to determine the weights of WL and WAL in the system. More theoretical studies are needed to fully understand the underlying physics. The resilience of the universal dependence to interface disorder should be a great advantage for future sophisticated multigate spintronic devices. Once the dependence of $\alpha_i$ on $\sigma$ is determined for a TFT structure with a constant a-IGZO channel quality at a constant $T$, the dependence should hold regardless of different strength of interface roughness scattering caused by different vertical electric fields imposed by gate voltages.  

The work was supported by the Ministry of Science and Technology of the Republic of China under Grant No. MOST 102-2112-M-003-009-MY3.

\end{document}